\documentstyle[aps,epsf]{revtex}

\begin{document}

\title{Brans-Dicke Cosmology with a scalar field potential}
\author{M. K. Mak\footnote{E-mail:mkmak@vtc.edu.hk} and T. Harko\footnote{E-mail: tcharko@hkusua.hku.hk}}
\address{Department of Physics, The University of Hong Kong, Pokfulam Road, Hong Kong, P. R. China.}
\maketitle

\begin{abstract}
Three solutions of the Brans-Dicke theory with a self-interacting quartic potential
and perfect fluid distribution are presented for a spatially flat FRW geometry.
The physical behavior is consistent with the recent cosmological scenario favored by type Ia supernova 
observations, indicating an accelerated expansion of the Universe.


\end{abstract}

\section{Introduction}

Recent observations of type Ia supernovae with redshift up to about 
$z\lesssim 1$ provided evidence that we may live in a low mass-density
Universe, with the contribution of the
non-relativistic matter (baryonic plus dark) to the total energy density of the Universe
of the order of $\ \Omega _{m}\sim 0.3$ \cite{1}. The value of  $\Omega _{m}$ is
significantly less than unity \cite{4}, and, consequently, either the Universe is
open or there is some additional energy density $\rho $ sufficient to reach
the value $\Omega _{total}=1$, predicted by inflationary theory. 
Observations also show that the deceleration parameter of the Universe $q$
is in the range $-1\leq q<0$, and the present-day Universe undergoes an
accelerated expansionary evolution.

Several physical models have been proposed to give a consistent physical
interpretation to these observational facts.
Photons propagating in extra-galactic magnetic fields can oscillate into very light
axions, so the supernovae appear dimer and more distant than they really are \cite{Cs02}.
For the missing
energy one candidate is the vacuum energy density, or the cosmological constant $\Lambda $ \cite{5}.
A charged Universe, in which the long range force carriers have a small mass, has been
considered as a cause for the cosmic acceleration in \cite{Br01}.
Another possibilities are cosmologies based on a mixture of cold dark matter
and quintessence, a slowly-varying, spatially inhomogeneous component \cite{6}.
An example of implementation of the idea of quintessence is the suggestion
that it is the energy associated with a scalar field $Q$ with
self-interaction potential $V(Q)$. If the potential energy density is
greater than the kinetic one, then the pressure associated to the $Q$-field is negative \cite{7}.
Quintessence models that accomodate the present day
acceleration tend to accelerate eternally, and, as a consequence, the resulting space-times
exhibit event horizons \cite{Fi01}. A wide class of quintessence models, with 
eternal acceleration, associated with static metrics, have been prezented in \cite{Go01}.

Brans-Dicke (BD) theory may explain the present
accelerated expansion of the Universe without resorting to a cosmological constant or
quintessence matter \cite{Ba01}. The conditions under which the dynamics of a self-interacting BD
field can account for the accelerated expansion have been considered in \cite{8} -\cite{11}.
Accelerated expanding solutions can be
obtained with a quadratic self-coupling of the BD field and a negative
coupling constant $\omega $ \cite{8}. A cosmic fluid obeying a perfect fluid
type equation of state cannot support the acceleration \cite{10}. The nature of the scalar field potential
compatible with a power law expansion in a self-interacting BD cosmology with a perfect 
fluid background has been analyzed in \cite{11}. Models with non-minimal
coupling of the scalar field have been considered in \cite{12}. Complicated cosmological scenarios,
with a four-dimensional effective action connected with supergravity and string theory, have been obtained in \cite{Se01}.

It is the purpose of the present Letter to consider some exact classes of solutions of
the field equations in the framework of BD theory with a quartic potential, $%
V(\phi )\sim \phi ^{4}$. By means of some appropriate transformations, the
field equations can be reduced to a system of two independent Riccati's type
differential equations. Three classes of exact solutions of the field
equations are presented, and their physical properties are investigated in
detail.

\section{Field Equations,Geometry and consequences}

The physical model we are considering is the Brans-Dicke action, along with a
self interacting potential $V\left( \phi \right) $, coupled to the matter
field Lagrangian $L_{m}$ via the action 
\begin{equation}\label{1}
S=\int d^{4}x\sqrt{-g}\left( \phi R-\frac{\omega }{\phi }\phi ^{,\alpha
}\phi _{,\alpha }-V\left( \phi \right) +L_{m}\right), 
\end{equation}
where $\omega $ is the BD coupling parameter. For $\omega =-1$ this action is identical
to the low energy string theory action. In the present Letter we use units so that $8\pi G=c=1$.

For a homogeneous flat
space-time, with scale factor $a$, filled with a
perfect fluid, with pressure $p_{m}$ and energy density $\rho
_{m}$, the Einstein-Brans-Dicke gravitational field equations are: 
\begin{equation}\label{2}
3H^{2}=\frac{\rho _{m}+\rho _{\phi }}{\phi },  2\dot{H}+3H^{2}=-\frac{%
p_{m}+p_{\phi }}{\phi },  
\end{equation}\label{3}
where $\rho _{\phi }$ and $p_{\phi }$ are the energy density and pressure
associated to the BD scalar field, given by 
\begin{equation}\label{31}
\rho _{\phi }\phi ^{-1}=\frac{\omega }{2}\psi ^{2}+\frac{V(\phi )}{2\phi }-3H\psi,
p_{\phi }\phi ^{-1}=\left( 1+\frac{\omega }{2}\right) \psi ^{2}-\frac{V(\phi )}{%
2\phi }+\dot{\psi}+2H\psi,  
\end{equation}
where we denoted  $H=\dot{a}a^{-1}$ and  $\psi =\dot{\phi}\phi ^{-1}$. 
$\psi $ gives the rate of change of the gravitational constant $G(t)$, $\psi =-\frac{\dot{G}(t)}{G(t)}$.

The wave equation for the BD field takes the form 
\begin{equation}\label{5}
\dot{\psi}+\psi ^{2}+3H\psi =\frac{\left( \rho _{m}-3p_{m}\right) \phi ^{-1}}{\left(
2\omega +3\right) }-\frac{1}{\left( 2\omega +3\right) }\left( \frac{2V(\phi )}{\phi 
}-\frac{dV(\phi )}{d\phi }\right).   
\end{equation}

The energy conservation of the matter implies $\dot{\rho _{m}}+3\left( \rho
_{m}+p_{m}\right) H=0$.

We assume that the self interaction potential $V(\phi )$ has the form 
$V\left( \phi \right) =V_{0}\phi ^{4}$, $V_0=constant$. This form corresponds to chaotic inflation model,
where the constant $V_{0}$ is subject to the constraint $V_{0}<10^{-12}$, coming from the
observational limits on the amplitude of fluctuations in the cosmic microwave background \cite{13}. 

With this choice, and with the use of Eqs.(\ref{2})-(\ref{5}), we obtain the following equation describing the dynamics
of the Universe:  
\begin{equation}\label{6}
\dot{\psi}-\frac{3}{\omega }\dot{H}=\frac{6}{\omega }H^{2}-\frac{1}{2}\psi
^{2}-3H\psi .  
\end{equation}

We introduce the following matrices 
\begin{equation}\label{7}
L=\left( \dot{\psi}-\frac{3}{\omega }\dot{H}\right) ,M^{T}=\left( 
\begin{array}{cc}
\psi  & H
\end{array}
\right) ,A=\left( 
\begin{array}{cc}
-\frac{1}{2} & -\frac{3}{2} \\ 
-\frac{3}{2} & +\frac{6}{\omega }
\end{array}
\right)  
\end{equation}

Then Eq.(\ref{6}) can be written as a matrix equation in the form $L=M^{T}AM$. 

We define two new variables $\hat{\psi},\hat{H}$, which are obtained
by means of a linear transformation, described by the matrix $K$, and
applied to the matrix $\hat{M}$ 
\begin{equation}\label{9}
M=\left( 
\begin{array}{c}
\psi  \\ 
H
\end{array}
\right) =K\hat{M}=\left( 
\begin{array}{ll}
f_{+}K_{+}^{-1} & f_{-}K_{-}^{-1} \\ 
K_{+}^{-1} & K_{-}^{-1}
\end{array}
\right) \left( 
\begin{array}{l}
\hat{\psi} \\ 
\hat{H}
\end{array}
\right) , 
\end{equation}
where $f_{+},f_{-},K_{+}$ and $K_{-}$ are real numbers.

In the new variables $\left( \hat{\psi},\hat{H}\right) $ we have $L=\hat{M}^{T}K^{T}AK\hat{M}$. We choose the elements of
the matrix $K$ so that $K^{T}AK$ is a diagonal matrix, with the the diagonal elements
$\lambda _{\mp }$ given by the eigenvalues of $A$. The eigenvectors and the
eigenvalues of the matrix $A$ are given by $\left( 
\begin{array}{cc}
f_{+}K_{+}^{-1} & K_{+}^{-1}
\end{array}
\right) ^{T}$, $\left( 
\begin{array}{cc}
f_{-}K_{-}^{-1} & K_{-}^{-1}
\end{array}
\right) ^{T}$ where $\lambda _{\mp }=\frac{12-\omega \mp s}{4\omega }$, $%
f_{\mp }=\frac{12+\omega \mp s}{6\omega }$, $K_{\mp }=\sqrt{f_{\mp }^{2}+1}$
and $s=\sqrt{144+24\omega +37\omega ^{2}}$. $s$ satisfies the condition $%
s\geq 0,\forall \omega $.

Therefore the equations satisfied by the
new unknown functions $\hat{\psi},\hat{H}$ are 
\begin{equation}\label{12}
a_{-}\frac{d\hat{\psi}}{dt}-\lambda _{-}\hat{\psi}^{2}=a_{+}\frac{d\hat{H}}{%
dt}+\lambda _{+}\hat{H}^{2}=F(t),  
\end{equation}
where we have introduced a solution generating function $F(t)$ and denoted $%
a_{\mp }=\varepsilon _{\pm }K_{\pm }^{-1}\left( f_{\pm }-3\omega
^{-1}\right) $, with $\varepsilon _{\pm }=\pm 1$.

\section{Classes of exact solutions of the field equations}

In order to obtain some classes of general solutions of Eq. (\ref{12}), we assume first that $%
\hat{\psi}$ and $\hat{H}$ are given by $\hat{\psi}=c_{-}t^{-1}$, $\hat{H}%
=c_{+}t^{-1}$, $c_{\pm }=const$. These functional forms of the new
variables satisfy Eq. (\ref{12}) if the consistency condition $\alpha _{-%
\text{ }}=\alpha _{+}$, relating the parameters $a_{\mp }$, $c_{\mp }$ and 
$\lambda _{\mp }$ holds, with $\alpha _{\mp }=-a_{\mp }c_{\mp }+\varepsilon
_{\mp }\lambda _{\mp }c_{\mp }^{2}$. By choosing $F(t)=\alpha
_{-}t^{-2}=\alpha _{+}t^{-2}$, we separate Eq. (\ref{12}) into a system of
Riccati's differential equations given as 
\begin{equation}\label{13}
a_{-}\frac{d\hat{\psi}}{dt}-\lambda _{-}\hat{\psi}^{2}=\frac{\alpha _{-}}{%
t^{2}},  a_{+}\frac{d\hat{H}}{dt}+\lambda _{+}\hat{H}^{2}=\frac{\alpha _{+}}{%
t^{2}}.  
\end{equation}

The mathematical form of $F(t)$ has been chosen in order to obtain an exact solution of Eqs. (\ref{12}).
The system (\ref{13}) has the particular solutions $\hat{\psi}_{0}=c_{-}t^{-1}$ and $%
\hat{H}_{0}=c_{+}t^{-1}$. With the help of the standard transformations $%
\hat{\psi}=u^{-1}+\hat{\psi}_{0}$, $\hat{H}=v^{-1}+\hat{H}_{0}$ the
Riccati's differential equations are transformed into two linear
Bernoulli's equations.

Therefore a first class of general solutions of Eqs. (\ref{13}) is given by
\begin{equation}\label{15}
\hat{\psi}=\frac{t^{2n_{-}}}{d_{-}-\frac{n_{-}}{c_{-}\left( 1+2n_{-}\right) }%
t^{2n_{-}+1}}+c_{-}t^{-1},\hat{H}=\frac{t^{-2n_{+}}}{d_{+}+\frac{n_{+}}{%
c_{+}\left( 1-2n_{+}\right) }t^{-2n_{+}+1}}+c_{+}t^{-1},  
\end{equation}
where we have denoted $n_{\mp }=c_{\mp }\lambda _{\mp }a_{\mp }^{-1}$. $%
d_{\mp }$ are arbitrary constants of integration.

With the use of the linear
transformation (\ref{9}) the solution for $\left( \psi ,H\right) $ is obtained in the form 
\begin{equation}\label{16}
\psi =\frac{f_{+}K_{+}^{-1}c_{-}+f_{-}K_{-}^{-1}c_{+}}{t}+f_{+}K_{+}^{-1}%
\frac{t^{2n_{-}}}{d_{-}-\frac{n_{-}}{c_{-}\left( 1+2n_{-}\right) }%
t^{2n_{-}+1}}+f_{-}K_{-}^{-1}\frac{t^{-2n_{+}}}{d_{+}+\frac{n_{+}}{%
c_{+}\left( 1-2n_{+}\right) }t^{-2n_{+}+1}},  
\end{equation}
\begin{equation}\label{17}
H=\frac{K_{+}^{-1}c_{-}+K_{-}^{-1}c_{+}}{t}+K_{+}^{-1}\frac{t^{2n_{-}}}{%
d_{-}-\frac{n_{-}}{c_{-}\left( 1+2n_{-}\right) }t^{2n_{-}+1}}+K_{-}^{-1}%
\frac{t^{-2n_{+}}}{d_{+}+\frac{n_{+}}{c_{+}\left( 1-2n_{+}\right) }%
t^{-2n_{+}+1}},  
\end{equation}

The BD scalar field and the scale factor of the Universe are: 
\begin{equation}\label{18}
\phi  =\phi _{0}t^{f_{+}K_{+}^{-1}c_{-}+f_{-}K_{-}^{-1}c_{+}}\left[ d_{-}-%
\frac{n_{-}}{c_{-}\left( 1+2n_{-}\right) }t^{2n_{-}+1}\right] ^{-\frac{%
f_{+}K_{+}^{-1}c_{-}}{n_{-}}}\left[ d_{+}+\frac{n_{+}}{c_{+}\left( 1-2n_{+}\right) }t^{-2n_{+}+1}\right]
^{\frac{f_{-}K_{-}^{-1}c_{+}}{n_{+}}},
\end{equation}
\begin{equation}\label{19}
a=a_{0}t^{K_{+}^{-1}c_{-}+K_{-}^{-1}c_{+}}\left[ d_{-}-\frac{n_{-}}{%
c_{-}\left( 1+2n_{-}\right) }t^{2n_{-}+1}\right] ^{-\frac{K_{+}^{-1}c_{-}}{%
n_{-}}}\left[ d_{+}+\frac{n_{+}}{c_{+}\left( 1-2n_{+}\right) }t^{-2n_{+}+1}%
\right] ^{\frac{K_{-}^{-1}c_{+}}{n_{+}}}, 
\end{equation}
where $a_{0}>0$ and $\phi _{0}>0$ are constants of integration.

Another class of exact solutions of the gravitational field equations in the
BD theory with matter fluid is obtained by assuming
for the generating function the form $F(t)=\beta =const.$. In this
case equations (\ref{12}) take the form 
\begin{equation}\label{a}
a_{-}\frac{d\hat{\psi}}{dt}=\lambda _{-}\hat{\psi}^{2}+\beta ,a_{+}\frac{d%
\hat{H}}{dt}=-\lambda _{+}\hat{H}^{2}+\beta .
\end{equation}

Depending on the sign of the parameters $\lambda _{-}$, $\lambda _{+}$ and $%
\beta $, we obtain two distinct classes of solutions. In the first case we
assume that $\lambda _{-}>0$, $-\lambda _{+}>0$ and $\beta <0$. With this
choice the general solutions of Eqs. (\ref{a}) are given by 
\begin{equation}
\hat{\psi}=-\sqrt{\left| \beta \right| \lambda _{-}^{-1}}\coth \left[
N_{-}\left( t-t_{0}\right) \right] ,\ \hat{H}=-\sqrt{-\left| \beta \right|
\lambda _{+}^{-1}}\coth \left[ N_{+}\left( t-t_{0}\right) \right] ,
\end{equation}
where $N_{\pm }=\sqrt{\epsilon _{\mp }\lambda _{\pm }\left| \beta \right| }%
a_{\pm }^{-1}$. This solution is mathematically consistent for values of $%
\hat{\psi}$ and $\hat{H}$ so that $\hat{\psi}>\sqrt{\frac{\left| \beta
\right| }{\lambda _{-}}},\hat{\psi}<-\sqrt{\frac{\left| \beta \right| }{%
\lambda _{-}}}$ and $\hat{H}>\sqrt{\frac{\left| \beta \right| }{-\lambda _{+}%
}}$.

With the use of the transformation (\ref{9}) we obtain 
\begin{equation}
\psi =-f_{+}K_{+}^{-1}\sqrt{\left| \beta \right| \lambda _{-}^{-1}}\coth %
\left[ N_{-}\left( t-t_{0}\right) \right] -f_{-}K_{-}^{-1}\sqrt{-\left|
\beta \right| \lambda _{+}^{-1}}\coth \left[ N_{+}\left( t-t_{0}\right) %
\right],
\end{equation}
and 
\begin{equation}
H=-K_{+}^{-1}\sqrt{\left| \beta \right| \lambda _{-}^{-1}}\coth \left[
N_{-}\left( t-t_{0}\right) \right] -K_{-}^{-1}\sqrt{-\left| \beta \right|
\lambda _{+}^{-1}}\coth \left[ N_{+}\left( t-t_{0}\right) \right]
\end{equation}

On integration, we obtain the BD scalar field and the scale factor: 
\begin{equation}
\phi =\phi _{0}\sinh ^{-\frac{f_{+}K_{+}^{-1}\sqrt{\left| \beta \right|
\lambda _{-}^{-1}}}{N_{-}}}\left[ N_{-}\left( t-t_{0}\right) \right] \sinh
^{-\frac{f_{-}K_{-}^{-1}\sqrt{-\left| \beta \right| \lambda _{+}^{-1}}}{N_{+}%
}}\left[ N_{+}\left( t-t_{0}\right) \right] ,
\end{equation}
\begin{equation}
a=a_{0}\sinh ^{-\frac{K_{+}^{-1}\sqrt{\left| \beta \right| \lambda _{-}^{-1}}%
}{N_{-}}}\left[ N_{-}\left( t-t_{0}\right) \right] \sinh ^{-\frac{K_{-}^{-1}%
\sqrt{-\left| \beta \right| \lambda _{+}^{-1}}}{N_{+}}}\left[ N_{+}\left(
t-t_{0}\right) \right] .
\end{equation}

A third class of solutions is obtained by assuming $\lambda _{-}>0$, $%
-\lambda _{+}>0$ and $\beta >0$. The solution can be formally obtained from
the previous one by means of the substitution $\beta \rightarrow -\beta $,
$\coth ix\rightarrow \frac{1}{i}\cot x$ etc.
In all these cases the energy density and pressure of the baryonic matter and of the
scalar field follow from the field equations (\ref{2})-(\ref{31}).

\section{Discussions and final remarks}

The first class of solutions depends on the set of five arbitrary constants $c_{-}$, $d_{\pm}$, $V_{0}$ and $\omega $.
To determine them from the actual observational data we take the matter pressure to be zero, $p_ {m}=0$. For the age of the
Universe and the Hubble constant we adopt the values $t=14Gyr$ and $H_0=65kms^{-1}Mpc^{-1}$ \cite{1}. The density parameters of the matter and of the BD scalar fields are defined as 
$\Omega _{m}=\frac{\rho _{m}}{3H^{2}\phi }=0.25$ and $\Omega _{\phi }=\frac{\rho
_{\phi }}{3H^{2}\phi }=0.75$, respectively \cite{1}.  
For the deceleration parameter $q=\frac{d}{dt}\frac{1}{H}-1$, which is an indicator of the accelerating behavior, we assume an actual value of $q=-0.5$.
Therefore, by taking $\omega $ as a free parameter, we have four observational
constraints to be satisfied by the model. Thus the numerical values of the constants $c_{-}$, $d_{\pm}$ and $V_{0}$ can be obtained from fitting the model with the observations.
Generally, a physical solution for the resulting non-linear system of algebraic equations
can be obtained only for small negative values of $\omega $.

The variation of the energy density of the matter and of the BD field is represented, for the first class of solutions, in Fig. 1. Generally, the energy density of the scalar field dominates the matter energy density,
thus providing the dominant contribution to the total energy density. The deceleration parameter
is represented, for different values of $\omega <0$, in Fig. 2. $q$ is negative,
with values in the range $-1<q<0$, indicating an accelerating evolution of the Universe. 

\begin{figure}
\epsfxsize=8cm
\centerline{\epsffile{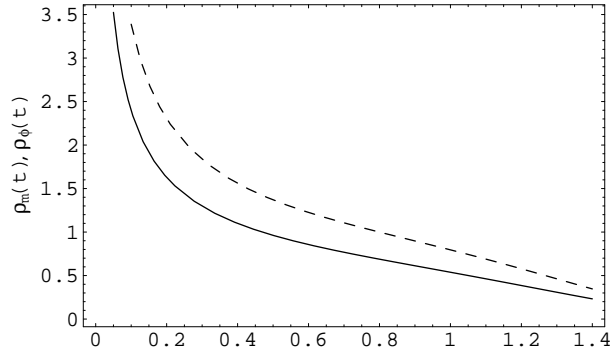}}
\caption{Time evolution of the energy density of the matter $\rho _{m}$ (solid curve) and of the scalar field $\rho _{\phi }$ (dashed curve) (in units of $10^{-47}GeV^4$)
for the first class of solutions 
for $\omega =-2$ ,$d_{-}=0.32$, $d_{+}=0.53$, $c_{-}=-1.95$, $V_0=0.0064$ and $c_{+}=-0.54$. The time is expressed in units of $10$ Gyr.}
\label{FIG1}
\end{figure}

\begin{figure}
\epsfxsize=8cm
\centerline{\epsffile{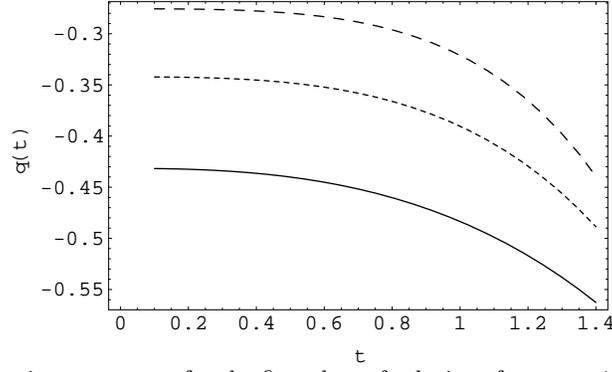}}
\caption{Dynamics of the deceleration parameter for the first class of solutions 
for $\omega =-1.8$ (solid curve),
for $\omega =-2$ (dotted curve),
and for $\omega =-2.2$ (dashed curve). We have used the values 
$d_{-}=0.32$, $d_{+}=0.53$, $c_{-}=-1.95$, $V_0=0.0064$ and $c_{+}=-0.54$. The time is expressed in units of $10$ Gyr.}
\label{FIG2}
\end{figure}

The second class of solutions depends (by chosing $t_0=0$) on three parameters $\beta $, $V_{0}$ and $\omega $.
Their numerical values can be obtained so that the solution fits the actual observational
values of $H$, $\Omega _{m}$ and $\Omega _{\phi }$. The resulting algebraic system of equations
has also a physical solution only for small negative values of $\omega $.

The time variations of the energy density of the matter and of the BD field and of $q$ for
the second class of solutions are represented in
Figs. 3 and 4. The energy density of the scalar field dominates the matter
energy density, and the evolution is accelerating.

\begin{figure}
\epsfxsize=8cm
\centerline{\epsffile{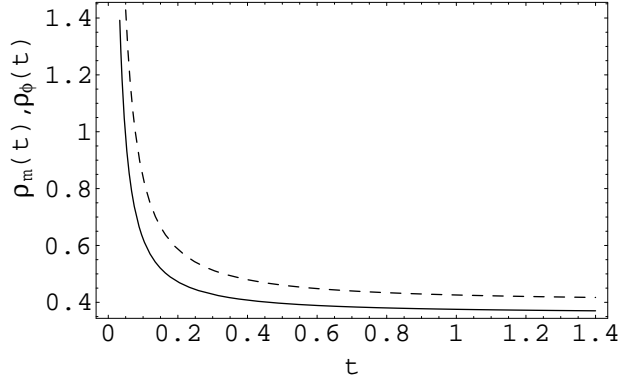}}
\caption{Time evolution of the energy density of the matter $\rho _{m}$ (solid curve) and of the scalar field $\rho _{\phi }$ (dashed curve) (in units of $10^{-47}GeV^4$)
for the second class of solutions, 
for $\omega =-2$, $\beta =0.001$ and $V_0=0.0073$. The time is expressed in units of $10$ Gyr.}
\label{FIG3}
\end{figure}

\begin{figure}
\epsfxsize=8cm
\centerline{\epsffile{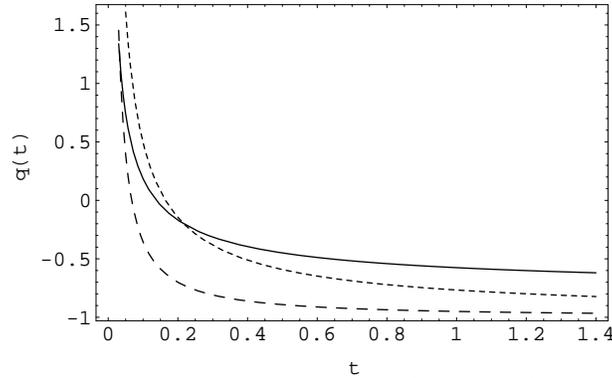}}
\caption{Dynamics of the deceleration parameter in the second class of solutions 
for $\omega =-1.8$ (solid curve),
$\omega =-2$ (dotted curve)
and $\omega =-2.2$ (dashed curve). In all cases we have used the values $\beta =0.001$ and $V_0=0.0073$. The time is expressed in units of $10$ Gyr.}
\label{FIG4}
\end{figure}

Therefore, these two classes of solutions are compatible with the observed
cosmological data only for small negative values of the coupling parameter
of the BD field. An other important observational bound requires $\Omega
_{\phi }<0.044$ at nucleosynthesis \cite{Be01}. This condition can also be
satisfied by appropriately chosing the matter equation
of state in the very early stages of evolution of the Universe. Generally,
in the present models we have $\rho _{\phi }>\rho _{m}$, and the evolution
is accelerating. An accelerating expansion during the whole evolution of the
Universe is not favored by the curent models of structure formation.
However, as shown in \cite{Pe02}, the scalar field, which is non-minimally
coupled to gravity, may undergo clustering processes, eventually forming
density perturbations, which can be investigated only within a non-linear
approach, and leading to the formation of some cosmic structures. On the
other hand, to satisfy the actual observational constraints, a much larger
value than that predicted by the inflationary scenario is needed for the
constant $V_{0}$.

\section*{Acknowledgments}

The authours would like to thank to the two anonymous referees, whose comments helped
to significantly improve the manuscript.

\end{document}